# Thickness dependent properties in oxide heterostructures driven by structurally induced metal-oxygen hybridization variations

*Zhaoliang Liao, Nicolas Gauquelin, Robert J. Green, Sebastian Macke, Julie Gonnissen, Sean Thomas, Zhicheng Zhong, Lin Li, Liang Si, Sandra Van Aert, Philipp Hansmann, Karsten Held, Jing Xia, Johan Verbeeck, Gustaaf Van Tendeloo, George A. Sawatzky, Gertjan Koster, Mark Huijben\*, Guus Rijnders*

Dr. Z.L. Liao, Dr. L. Li, Prof. M. Huijben, Prof. G. Koster, Prof. G. Rijnders
MESA+ Institute for Nanotechnology, University of Twente, 7500 AE, Enschede, The Netherlands
E-mail: m.huijben@utwente.nl
Dr. N. Gauquelin, J. Gonissen, Prof. S. van Aert, Prof. J. Verbeeck, Prof. G. van Tendeloo
Electron Microscopy for Materials Science (EMAT), University of Antwerp, 2020 Antwerp, Belgium
Dr. R.J. Green, Dr. S. Macke, Prof. G.A. Sawatzky
Quantum Matter Institute and Department of Physics and Astronomy, University of British Columbia, 2355 East Mall, Vancouver, V6T 1Z4, Canada
Dr. S. Thomas, Prof. J. Xia
Department of Physics and Astronomy, University of California, Irvine, Irvine, CA, 92697, USA
Dr. Z. Zhong, Dr. L. Si, Prof. K. Held
Institute of Solid State Physics, TU WIEN, A-1040 Vienna, Austria
Dr. S. Macke, Dr. Z. Zhong, Dr. P. Hansmann
Max Planck Institute for Solid State Research, Heisenbergstraße 1, 70569 Stuttgart, Germany
Dr. R.J. Green
Max Planck Institute for Chemical Physics of Solids, Nöthnitzerstraße 40, 01187 Dresden, Germany

**Keywords**: metal-oxygen hybridization, oxygen octahedral coupling, manganites, interface, oxide heterostructure

**Thickness driven electronic phase transitions are broadly observed in different types of functional perovskite heterostructures. However, uncertainty remains whether these effects are solely due to spatial confinement, broken symmetry or rather to a change of structure with varying film thickness. Here, we present direct evidence for the relaxation of oxygen 2p and Mn 3d orbital (p-d) hybridization coupled to the layer dependent octahedral tilts within a La$_{2/3}$Sr$_{1/3}$MnO$_3$ film driven by interfacial octahedral coupling. An enhanced Curie temperature is achieved by reducing the octahedral tilting via interface structure engineering. Atomically resolved lattice, electronic and magnetic structures together with X-ray absorption spectroscopy demonstrate the central role of**

thickness dependent p-d hybridization in the widely observed dimensionality effects present in correlated oxide heterostructures.

## 1. Introduction

A diverse range of phenomena spanning from colossal magnetoresistance to metal-to-insulator transition (MIT) and superconductivity have been attracting tremendous interest in transition metal oxides (TMOs) for both fundamental research as well as technological applications.[1] Of particular interest are the phase transitions in TMOs, which can be manipulated by tuning either bandwidth or band-filling.[2,3] Various phase transitions (MIT, ferroelectricity to paraelectricity, ferromagnetic to antiferromagnetic) can already be achieved in different types of TMO heterostructures just by varying the thickness.[4-10] In recent literature, several physical models are used as a design principle in order to achieve control over the reconfigurable properties by changing the thickness. The reduced dimensionality tipping the delicate competition between collective quantum phases which are proposed to contribute to the MIT in ultrathin $LaNiO_3$ and $SrVO_3$ thin films[4-6] are very well-known examples of thickness effects on properties, while a polar catastrophe is proposed to generate a magnetic phase transition in polar $LaMnO_3$ ultrathin films.[7]

The recently revealed nontrivial oxygen octahedral coupling at perovskite oxide interface which induces the modulation of oxygen coordination across the interface should play an important role in affecting the delicate spin-charge-orbital coupling in those correlated heterostructure systems.[11-15] Scanning transmission electron microscopy (STEM) using the latest scanning and imaging techniques for both $SrRuO_3$[12] and $La_{2/3}Sr_{1/3}MnO_3$ (LSMO)[13] clearly show a thickness dependent oxygen coordination and resultant modified spin-orbital coupling. Our previous study further suggests that a reduced octahedral distortion is corresponding to an increased Curie temperature in LSMO films.[13] Given the close similarity of the structures in the previously mentioned examples[4-10] with these latter systems,[12,13] the question arises whether a similar structural effect could be essential in the former studies and how is the structure coupled to spin-charge-orbital degrees of freedom in giving rise to those emergent phenomena. The missing link is an exact physical relationship between the observed modified structure, the changes of electronic structure and the underlying physical phenomena for these thickness driven phase transitions. Here, we provide experimental evidence for the structure-property relationship in an epitaxial model system, LSMO on $NdGaO_3$ (NGO), where orbital hybridization is identified as the main driving mechanism for thickness dependent property changes. By reducing the octahedral tilting and hence increasing the p-d hybridization via interfacial structure engineering, ferromagnetism can be stabilized and an increased Curie temperature is achieved. It is very plausible that similar effects play a role in other oxide heterostructures and can be used as a leading design principle crucial for further development of these functional heterostructures into miniaturized oxide electronic devices.

The half metallic ferromagnet LSMO is of great importance due to its 100% spin polarized nature[16,17] and is a promising material for oxide spintronic applications. For LSMO thin films previous studies have already shown that the ferromagnetic behavior severely depends on the film thickness.[10,18-24] Tremendous efforts have been made to unravel this thickness dependence, but its origin is still debated.[10,18-27] The variation of $Mn^{3+}/Mn^{4+}$ ratio, orbital reconstruction due to strain and/or broken symmetry together with extrinsic structure imperfection due to growth are several proposed mechanisms responsible for this deterioration of ferromagnetism.[18-25] Preference of either $3z^2-r^2$ or $x^2-y^2$ occupation is suggested to favor an antiferromagnetic ground state,[18,19,28,29] while the enrichment of $Mn^{3+}$ would also enhance Jahn-Teller distortions.[23,24] However, an interfacial charge transfer limited to ~0.13 eV and confined to ~ 2 unit cells (uc)[30] cannot explain the long range thickness dependence in thicker films, although crucial for local magnetism.[25] Also the observed ferromagnetic behavior of $La_{1-x}Sr_xMnO_3$ (with $T_C>250$ K) over a wide range of compositions $0.15<x<0.5$[16] indicates the limited impact of any possible charge transfer on the Curie temperature. A roughly ~3 uc length scale of charge leakage as demonstrated in $(LaMnO_3)_{2n}/(SrMnO_3)_n$ superlattice should further suppress the impact of charge transfer.[31] In addition, possible changes in the overlap of the orbitals due to structural changes were disregarded up to now, although a strong correlation with magnetic behavior has been demonstrated in bulk manganites.[3,32,33] Recently it has been shown that the oxygen octahedral rotation pattern in LSMO films can be tuned by direct coupling to the oxygen octahedral rotation pattern of the underlying substrate.[13] NGO substrates provide a specific octahedral tilt angle, and corresponding orbital overlap, which can be reduced by an increased thickness of the LSMO layer as well as by incorporation of a $SrTiO_3$ (STO) buffer layer. In the present study we will unravel the correlation between the orbital overlap and the thickness dependence by systematically investigating the relation between magnetic behavior, crystal structure, orbital reconstruction and bandwidth in LSMO films of various thicknesses using buffered and capped samples. Our results demonstrate the dominating role of the local O-2p and Mn-3d (p-d) orbital hybridization to the ferromagnetic ordering at the atomic scale in LSMO films and they reveal that spatial variation in p-d orbital hybridization, due to octahedral relaxation within LSMO, is responsible for the observed thickness dependence in material properties.

## 2. Results and discussion

LSMO thin films were grown by pulsed laser deposition on $NdGaO_3$ (110) (NGO) substrates (see Methods). Control over the near-interfacial crystal structure was achieved by atomically precise thickness control and the use of ultrathin buffer and capping layers, or combinations thereof, which will hereafter be referred to as *BL* and *CL* respectively,[13] thereby inducing the hypothesized variation of the orbital overlap in LSMO films. The oxygen octahedral rotation pattern across a 20 uc LSMO/NGO ($L_{20}$/NGO) interface was imaged by annular bright field STEM (ABF-STEM) (see **Figure 1a**). The $MnO_6$ octahedra are shown to follow the tilt angle of NGO in the interface region due to the oxygen octahedral coupling (OOC) effect,[11,13] which progressively decays away from the interface and is already negligible in the 4$^{th}$ uc. Figure 1b shows the profile of projected B-O-B bond angle $\Theta$ in (001) plane, which is determined using statistical parameter estimation theory[34,35] (See Supplementary Figure S1). The bond angle $\Theta$ in LSMO right at the interface is comparable to the angle in the NGO substrate (= 165°) and increases, starting from the interface, up to a saturation value of ~ 173° above the 4$^{th}$ LSMO uc. Further away from the interface, a long range slow relaxation tail of the octahedral rotation pattern should exist, similar to LSMO/STO heterostructures.[26]

The existence of a slow relaxation tail is evidenced by an increasing out-of-plane lattice parameter *c* with increasing film thickness, as shown in Figure 1c (for the measurement of *c*, see Supplementary Figure S2). Additionally, upon the introduction of a STO *BL*, which renders the octahedral rotation pattern in the LSMO film to behave like bulk already starting from the interface (See Ref. [13] and Supplementary Figure S3), the lattice constant *c* is increased (see Figure 1c). For example, the increased lattice constant *c* of a $L_{30}/S_6$ film becomes almost identical to that of a non-buffered 90 uc LSMO ($L_{90}$). Note that the buffered LSMO films possess the same in-plane lattice constant as the non-buffered LSMO films, as confirmed by the reciprocal space mapping (see Supplementary Figure S2). On the other hand, introducing a STO *CL* does not change the LSMO lattice constant (see Figure 1c), in contrast to a STO *BL*. From the observed structural effects, we can infer that an increase of the lattice constant *c* due to increasing thickness or induced by a STO *BL* seems to suggest that the presence of a long range structural relaxation is more likely caused by interface OOC rather than strain accommodation.

The relationship between structure and measured magnetic properties is investigated by using the variation of $T_C$ for films with different thicknesses of LSMO and presence/absence of a STO *BL/CL* as shown in Figure 1d. Parallel to the octahedral rotation happening at the direct proximity to the interface in LSMO discussed above, $T_C$ is dramatically reduced in ultrathin

LSMO, e.g., down to 50 K for a 4 uc LSMO film (see Figure 1d). With increasing thickness without variation of the OOC effect, the $T_C$ increases and is promoted to its bulk value of 350 K in relatively thick films (> 30 uc).

Interestingly, films with a larger out-of-plane lattice constant have a higher $T_C$ (see Figure 1c-d). In accordance, the increased lattice constant *c* of a 30 uc LSMO film by a STO *BL* corroborates with a $T_C$ increase from 332 K to 350 K. In contrast, the presence of a STO *CL* does not influence the $T_C$ ($L_{30}$ vs. $S_9/L_{30}$), where it was already observed that the structure of the LSMO was not affected. The difference between the effect of *BL* and *CL* on magnetism again suggests a pivotal role of octahedral tilt in the magnetism of the LSMO films. A record recovery of $T_C$ by ~ 120 K was observed for buffered 6 uc LSMO films ($L_6/S_{23}$) (see Figure 1d), consistent with the significant reduction of the octahedral tilt in LSMO by such STO *BL*.[13]

A more detailed investigation of the role of the STO *BL* on magnetization and transport is shown in **Figure 2**. In addition to the enhancement of $T_C$, the saturated magnetic moment ($M_S$) is increased when a STO *BL* is present (see Figure 2a). Reduced saturated magnetism in the $L_6$ film compared to bulk (e.g., $L_{90}$) is mainly due to the presence of a structural distortion near the interface and the existence of a surface magnetic dead layer.[13] This surface dead layer can be reduced by introducing a STO *CL*. Figure 2a shows that both the non-buffered and buffered 6 uc LSMO films exhibit a higher saturated magnetic moment after the introduction of a 2 uc STO *CL*. For the $S_2/L_6/S_9$ film, the $M_S$ is equal to 3.4 $\mu_B$/Mn at 25 K, close to the value (~3.7 $\mu_B$/Mn) in bulk films,[36] indicating the absence of a magnetic dead layer. The *CL* significantly increases the $T_C$ of non-buffered ultrathin LSMO films (e.g., $S_2/L_6$) whereas it has just a minor impact on the $T_C$ of buffered LSMO films (e.g., $S_2/L_6/S_9$) (see Figure 2a) and almost no effect on the $T_C$ of thick LSMO (e.g., $S_9/L_{30}$) (see Figure 1d). This indicates a very local effect of the *CL* on the topmost layers of the films. The increased magnetism with increasing thickness of the *BL* suggests a gradual structure relaxation within the STO layers. In addition to this stabilized magnetism, the buffered LSMO films exhibit a higher conductivity (see Figure 2b). These modifications of magnetic and electrical properties, through capping, buffering and thickness variations, point to the importance of structure relaxation in these thickness driven phenomena.

To ensure that this enhancement of ferromagnetism is exclusively caused by the presence of a STO *BL/CL*, a 6 uc LSMO sample with a patterned STO *BL* and *CL* is fabricated (See Figure 2c for schematic; for the STO patterning-process, see Supplementary Figure S4). Left- and right-circularly polarized light is focused on the LSMO layer and reflected. The phase difference between the two polarizations is measured using a scanning Sagnac Interferometer.[37] This enables the mapping of the Kerr intensity, indicating three distinguishable regions with different magnetic properties (see Figure 2c). In the Kerr map, the Curie temperatures for the different regions were determined from the temperature dependent Kerr signal ($\theta_K$) as shown in Figure 2d. $T_C$ is determined from the second derivative of the Kerr signal (see Supplementary Figure S5). $T_C$ for region A ($S_2/L_6$), B ($L_6$) and C ($L_6/S_3$) are 191 K, 135 K and 213 K respectively, which are quite consistent to values determined by vibrating sample magnetometry (VSM) shown in Figure 1d. This experimental result unambiguously demonstrates the enhancement of ferromagnetism due to STO *BL* and *CL*, fully excluding sample to sample variation.

We now turn to the vertical distribution of magnetism in the layers. For that purpose we determined a depth resolved magnetic profile probed by synchrotron resonant soft x-ray reflectivity (RSXR)[38,39] (see Supplementary Figure S6). Without *CL*, both the buffered and non-buffered 6 uc LSMO films showed a magnetic dead layer near the top surface of LSMO layers (see **Figure 3a-b**). This suggests that the STO *BL* apparently is able to increase the overall Curie temperature, but cannot remove the surface dead layer. Yet the surface dead layer was absent when the sample was capped with STO, both for the buffered and non-buffered LSMO films, see Figure 3c-d. This enhanced magnetism due to the STO *CL* revealed by RSXR was consistent with the earlier described magnetization measurement using VSM (see Figure 2a). From the RSXR one can extract a stoichiometry profile for both buffered and non-buffered LSMO films, verifying their similarities and sharp chemical contrast across interfaces. RSXR also reveals a stoichiometric profile from interface to bulk region except the top surface. The variation of the La/Sr stoichiometry near the top surface is probably important to explain the role of the *CL*, since the capped samples have a more stoichiometric surface than non-capped samples (see Figure 3). An enhanced surface structural distortion, as indicated by our DFT calculation (see Supplementary information), may also play a strong role in inducing a surface dead layer, while a STO *CL* will recover surface magnetism by reducing the surface distortion, similar to observations for $LaNiO_3$ films.[40]

Having established a strong connection of the magnetism to the local structure at or near heterointerfaces, we now turn to investigate any role that orbital reconstruction might play. Based on an orbital reconstruction model,[18,19,28] a stronger $3z^2-r^2$ orbital reconstruction is expected with an increased *c/a* ratio which favors C-type anti-ferromagnetism, degrading ferromagnetism in $L_{30}/S_6$ films as compared to $L_{30}$ films. To quantify how the orbital reconstruction could influence the magnetism in LSMO, the Mn $3z^2-r^2$ and $x^2-y^2$ orbitals were preferentially probed by in-plane (σ) and out-of-plane (π) polarized X-ray absorption spectroscopy (XAS) respectively at the Mn $L_{2,3}$ edge (see **Figure 4a**). The measured XAS is shown in top panel of Figure 4b. For different films, their two spectra $I_σ$ and $I_π$ all show slight difference, indicating a small linear dichroism. The X-ray linear dichroism (XLD) is calculated as the intensity difference between the signals with different polarization ($I_σ - I_π$). As shown in the bottom panel of Figure 4b, all films exhibit quite similar XLD characteristics, e.g., $L_6$ and $L_6/S_9$ on the one hand; $L_{30}$ and $L_{30}/S_6$ on the other hand. The relative $3z^2-r^2$ to $x^2-y^2$ orbital occupancy can be quantified using sum-rules of the XLD (ALD) spectra. A positive (or negative) ALD indicates a preferential occupation of $3z^2-r^2$ (or $x^2-y^2$) orbital.[20] By integrating the spectra in the range 649-659 eV (around the Mn $L_2$ edge), the area under the XLD peak is obtained and plotted as a function of $T_C$ for different films (see Figure 4c). The ALDs of all different films are ~1%, indicating a very small orbital polarization with light preferential occupation of the $3z^2$-$r^2$ orbital. Since the XLDs were taken at 300 K, well above the $T_C$ of LSMO films thinner than 30 uc (Figure 4c), therefore, the contribution of magnetization to XLD can be ruled out. A change of orbital polarization is not accompanied by any change of the Curie temperature. Therefore one can infer that the small degree of orbital polarization present plays a minor role in magnetic ordering when the lattice mismatch is relatively small.

It is noteworthy that the 6 uc LSMO has a smaller orbital polarization than thicker films. This indicates a competitive action of compressive strain and octahedral tilt on tetragonal distortions of the octahedra. Since the Mn-Mn distances in the (110) plane are locked-in by the NGO substrate, the tilting of the octahedra will increase the Mn-O bond length in the (110) plane, compensating the effect from compressive strain which tends to elongate the Mn-O bond in the direction normal to the interface. This effect subsides with increasing thickness, and hence orbital polarization will increase in thicker films. With a STO *BL*, the octahedral tilt disappears in ultrathin LSMO,[13] squeezing the in-plane Mn-O bond, resulting in an increased occupation of the $3z^2$-$r^2$ orbital in $L_6/S_9$ film as shown in Figure 4c. The STO *CL* is also found to enhance orbital polarization of both non-buffered and buffered 6 uc LSMO, indicating a bulk like electronic structure at the surface layer, which can be connected to the recovery of surface magnetism as illustrated above by RSXR (see Figure 3). Although a large orbital polarization has been found to favor an antiferromagnetic insulating phase in largely strained LSMO/DyScO$_3$[41] and LSMO/LaAlO$_3$ films,[42] the orbital polarization in the LSMO films on NGO substrates seems to be so small that it plays only a minor role for the ferromagnetic ordering. Furthermore, the XAS results of different LSMO films (Figure 4b) show identical peak position of the Mn $L_{2,3}$ edge, indicating that a valence change doesn't play a significant role as well. Our previous cross-sectional STEM characterization and X-ray reciprocal space mapping also demonstrate an unchanged orthorhombic symmetry for different thicknesses of LSMO films on NGO (110).[13,14] Therefore, local bonding strength should play a central role and the ferromagnetic ordering of LSMO films grown on NGO is more likely determined by p-d hybridization.[33]

In fact, the relaxation of octahedral tilts can be connected to a spatial variation of p-d hybridization within LSMO as schematically illustrated in Fig. 5a. With larger octahedral tilt, the overlap of the O-2p and Mn-3d orbitals is smaller. Due to the OOC effect, the Mn-O-Mn bond angle near the interface is close to the substrate Ga-O-Ga bond angle of (154°) as discussed above. This corresponds to the critical bond angle for the paramagnetic to ferromagnetic phase transition in $La_{2/3}R_{1/3}MnO_3$.[33] Furthermore, the tilting of the octahedra together with the compressive strain effect will elongate both the in-plane and out-of-plane Mn-O bonds as mentioned above, and thus the octahedral tilt and elongated Mn-O bond will cooperatively reduce the p-d hybridization and bandwidth as is demonstrated by first principle density functional theory calculations (see supplementary Table S1 and Figure S7). Our DFT calculations indicate that the bandwidth of the $e_g$ orbital is reduced by 28% due to a decrease of the Mn-O-Mn bond angle from 180° to 151.5° while maintaining the Mn-O bond length. The reduction of bandwidth can be further promoted to 32% if the Mn-Mn distance is kept constant when changing the bond angle from 180° to 151.5°.

Experimentally, the effect of the octahedral tilt on the strength of p-d hybridization is directly measured by atomically resolved fine-structure changes at the O-K edge in electron energy loss spectroscopy (EELS). **Figure 5b** shows layer resolved spectra of the O-K edge spectrum taken from the individual layers across a non-buffered LSMO film. There is a strong peak in the pre-edge region (528-531 eV), revealing that the holes introduced by Sr doping have a strong mixture of O-2p and Mn-3d character.[43,44] When getting closer to the LSMO/NGO interface, the pre-peak intensity is found to decrease within the first 6 uc (see inset panel in Figure 5b), suggesting a weaker O-2p and Mn-3d hybridization.[45,46] Although more occupancy of Mn $e_g$ orbital will lead to a reduction of pre-edge intensity, it is not very significant I($Mn^{3+}$)/I($Mn^{4+}$) = 0.8. Furthermore, a sole valence effect is expected to induce a slightly higher intensity near the interface due to a small increase in hole doping of interfacial Mn. Therefore, the reduced pre-edge intensity arises from a structure variation (see also supplementary Figure S8). The change of the orbital hybridization from the bulk region towards the interface agrees well with the schematic structure profile shown in Figure 5a. By introducing a STO *BL*, which has the effect of suppressing the $MnO_6$ octahedral tilt near the interface, the O-K edge becomes identical at the near-interface region than in bulk (see Figure 5c). We can conclude that a reduction of the p-d hybridization, which is a consequence of a change of the tilt of the $MnO_6$ octahedra, is at the origin of the reduction of $T_C$ with decreasing film thickness.

The effect of thickness on the electronic structure of LSMO is further investigated by O K-edge XAS analysis of a series of buffered and non-buffered films of various thickness. The pre-edge regions of the O K-edge of these different samples are shown in Figure 5d (For full spectra, see Supplementary Figure S9). At 50 K, three peaks at 529.5 eV, 530 eV and 532 eV are observed in 90 uc LSMO film ($L_{90}$) and assigned to $e_g\uparrow$, $t_{2g}\downarrow$ and $e_g\downarrow$ respectively.[43] The $L_{90}$ shows a spectrum quite similar to bulk crystalline LSMO,[43] indicating a bulk-dominated p-d hybridization in thick LSMO. For $L_{30}$ film, which has lower $T_C$ than bulk (see Figure 1d), the pre-edge has a lower intensity. However, a 6 uc STO *BL* can drive the pre-edge of 30 uc LSMO ($L_{30}/S_6$) to be similar to the one observed for a $L_{90}$ film. As a consequence, the $L_{30}/S_6$ film behaves identically to a $L_{90}$ film, e.g. same Curie temperature (see Figure 2a). For thinner 10 uc and 6 uc LSMO films, the reduced amount of absorption in the film relative to the substrate also reduces the pre-edge intensity, so the change of intensity cannot be directly interpreted for p-d hybridization. However, the different line shapes of the pre-edge still reflect a profound change of the Mn-O covalency. The most remarkable characteristic change of line shapes is the disappearance of the double-peak aspect of the pre-edge with reducing film thickness. The disappearance of this double peak is attributed to a loss of intensity of the first peak located at 529.5 eV ($e_g\uparrow$).[48,49] This is further suggested by a dip around the first peak in the $I(L_{30})$-$I(L_{30}/S_6)$ curve (see Figure 5d). The intensity of the third peak located at 530 eV ($e_g\downarrow$) is reduced as well as the thickness is decreased. Reduced intensity of $e_g\uparrow$ and $e_g\downarrow$ peaks has been interpreted as a signature of electron localization and more ionic environment of Mn, resulting in less hybridization between O and Mn.[46,49] Moreover, the double peak feature is recovered in buffered 6 uc LSMO film ($L_6/S_9$) (see Figure 5d). A spectrum from a pure STO crystal verifies that the STO buffer layer does not contribute to the prepeak region and only adds a bump at higher energies. This fact underlines the important role of structure distortion in p-d hybridization.

## 3. Conclusion

In conclusion, octahedral relaxation leading to a change of p-d hybridization driven by interfacial OOC appears to be the strongest factor in thickness related variations of magnetic and transport properties in epitaxial LSMO films on NGO substrates. Our study demonstrates that the reconstructed octahedral rotation pattern due to the OOC effect at the interface is a nontrivial factor of perovskite oxide interfaces and causes a thickness dependent orbital hybridization effect. Although LSMO has been used as a prototype material for this study, the results presented here can be applied to understand thickness related properties variations in many other correlated oxide perovskite systems and have deep implications for understanding emergent functionalities in those complex systems. Our results further reaffirm the important role of mixed transition metal 3d and oxygen 2p character in TMOs[2] and address the advantage of heterostructures whose interfaces allow us to freely control p-d orbital hybridization in the prospect of controlling and tailoring functionalities. The octahedral relaxation can be modified or even suppressed by engineering the interface octahedral rotation network, e.g. inserting a non-tilted STO buffer layer, which has the effect of enhancing magnetism. The magnetic dead layer can be eliminated by introducing STO capping layers, making LSMO very promising for near-future spintronic applications. The strong impact of a STO buffer layer on the properties of both ultrathin and thick LSMO films indicate an efficient way to engineer functionalities via local control of OOC. The capability to pattern such buffer and capping layers, as demonstrated here, also allows us to develop new types of devices with smaller length scales and new type of junctions. The revealed origin of thickness-properties correlation discussed here paves the way toward scaling down of correlated oxide systems in order to achieve smaller scale new full oxide electronics.

## 4. Experimental methods

The STO and LSMO layers were grown by pulsed laser deposition (PLD) at 680 $^{o}$C on atomically flat NGO (110) substrates from a single crystalline STO target and a stoichiometric La$_{2/3}$Sr$_{1/3}$MnO$_3$ ceramic target, respectively. The PLD used a KrF excimer laser operating at 248 nm and at a repetition rate of 2 Hz. The atomically flat NGO substrates were obtained by buffered HF chemical etching and subsequent annealing at 1050 $^{o}$C for 4 hours.[50] The oxygen partial pressure during growth was 0.2 mbar. The laser energy fluences for the growth of LSMO and STO were 0.6 J/cm$^2$ and 1 J/cm$^2$ respectively. The growth process was monitored by reflection high-energy electron diffraction (RHEED), which confirmed the layer by layer characteristic growth. The surfaces of the films were atomically flat as confirmed by atomic force microscopy. For patterned LSMO sample, the patterned STO layers were fabricated by using a shadow mask during growth, while there was no shadow mask for the growth of LSMO. This growth process ensured that the LSMO at different regions experienced same condition.

Transport and magnetism were characterized by Quantum Design Physics Properties Measurement System (QD-PPMS) and Vibrating Sample Magnetometer (VSM) respectively. The magnetization was obtained by subtracting the NGO paramagnetic signal.[13] The lattice parameters of the thin films were characterized by PANalytical-X'Pert Materials Research Diffractometers (MRD) at high resolution mode. Atomic scale characterization of the lattice structure using ABF-STEM imaging at 300 kV and electron energy loss spectroscopy (EELS) were performed on the Quant-EM instrument at the University of Antwerp. For STEM-EELS experiments, the beam was monochromated to achieve a 100 meV energy resolution and the microscope operated at 120 kV to reduce beam-damage. Cross-sectional cuts of the samples along the NGO [1-10] direction were prepared using a FEI Helios 650 dual-beam Focused Ion Beam device.

Stoichiometry and magnetic profiles were probed by using resonant soft X-ray reflectivity (RSXR). The RSXR experiments were performed using an in-vacuum 4-circle diffractometer at the Resonant Elastic and Inelastic X-ray Scattering (REIXS) beamline at Canadian Light Source (CLS) in Saskatoon, Canada. The beamline has a flux of $5 \times 10^{12}$ photon/s and photon energy resolution $\Delta E/E$ of $\sim 10^{-4}$. The base pressure of the diffractometer chamber was kept lower than $10^{-9}$ Torr. The samples were aligned with their surface normal in the scattering plane and measured at a temperature of 20 K. The measurements were carried out in the specular reflection geometry with several non-resonant photon energies as well as energies at the Mn $L_{2,3}$ resonance (~635-660 eV). X-Ray Linear Dichroism (XLD) was performed to investigate the 3d $x^2-y^2$ and $3z^2-r^2$ orbital occupancy by using σ polarized (polarization vector near perpendicular to the surface) and π polarized (polarization vector in the surface) photons. X-ray absorption spectroscopy (XAS) of oxygen K-edge was measured with the photon polarized along [001] axis of NGO.

The mapping of polar Kerr signal with spatial resolution of 2 μm was performed on a scanning Sagnac interferometer at University of California, Irvine, United States. A magnetic field of 0.2 T was applied along surface normal direction to cant the magnetization along out-of-plane direction. The Curie temperature was defined as the point in the Kerr signal versus temperature of maximum curvature.

**Acknowledgements**
M.H., G.K. and G.R. acknowledge funding from DESCO program of the Dutch Foundation for Fundamental Research on Matter (FOM) with financial support from the Netherlands Organization for Scientific Research (NWO). This work was funded by the European Union Council under the 7th Framework Program (FP7) grant nr NMP3-LA-2010-246102 IFOX. J.V. and S.V.A. acknowledge financial support from the Research Foundation Flanders (FWO, Belgium) through project fundings (G.0044.13N, G.0374.13N, G.0368.15N, G.0369.15N). The Qu-Ant-EM microscope was partly funded by the Hercules fund from the Flemish Government. N.G. acknowledges funding from the European Research Council under the 7th Framework Program (FP7), ERC Starting Grant 278510 VORTEX. N.G., J.G., S.V.A., J.V. acknowledge financial support from the European Union under the Seventh Framework Program under a contract for an Integrated Infrastructure Initiative (Reference No. 312483-ESTEEM2). The Canadian work was supported by NSERC and the Max Planck-UBC Centre for Quantum Materials. Some experiments for this work were performed at the Canadian Light Source, which is funded by the Canada Foundation for Innovation, NSERC, the National Research Council of Canada, the Canadian Institutes of Health Research, the Government of Saskatchewan, Western Economic Diversification Canada, and the University of Saskatchewan.

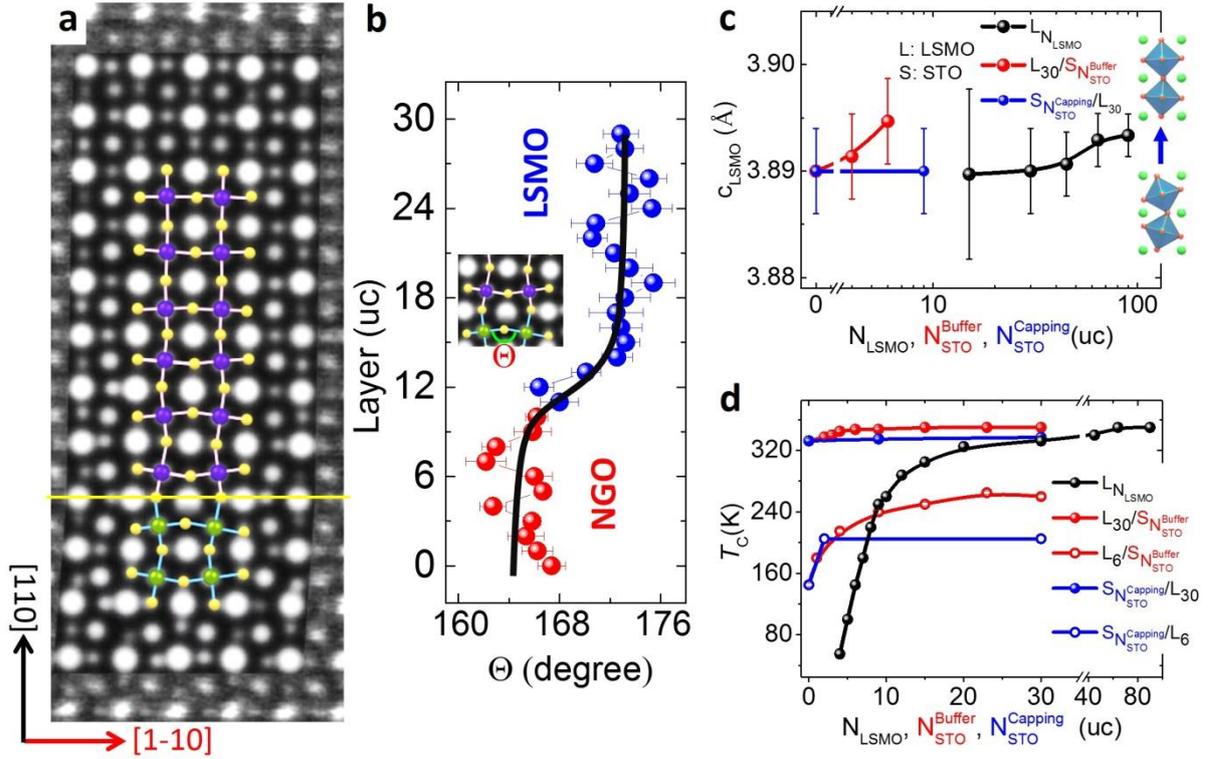

**Figure 1: Relaxation of octahedra induced thickness-properties correlation**. **a**, ABF-STEM image of LSMO/NGO cross-section. The intensity (I) is rescaled to $-I^{1/4}$ for better oxygen contrast. Inset is the refined parametric model using statistical parameter estimation. The colorized atoms and bonds highlight the relaxation of octahedral and B-O-B bond angle. **b**, The layer position dependent [1-10] directional and (001) plane projected B-O-B bond angle $\Theta$. Inset shows the definition of $\Theta$. **c,d**, The LSMO out-of-plane lattice constant and Curie temperature $T_C$ of LSMO films as a function of thickness of LSMO ($N_{LSMO}$), STO buffer layer ($N_{STO}^{Buffer}$) and capping layer ($N_{STO}^{Capping}$).

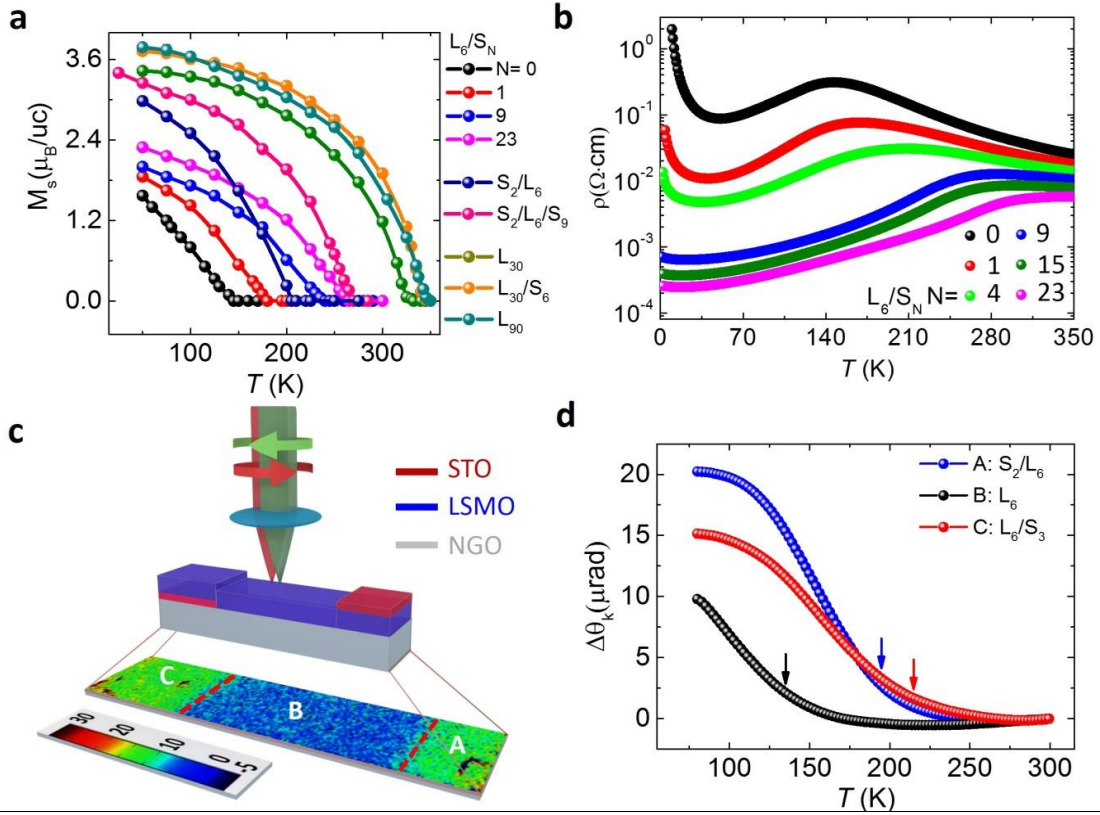

**Figure 2**: **Impact of STO buffer and capping layers on magnetic and transport properties of LSMO films. a**, Temperature dependent saturated magnetization for different thicknesses of LSMO films with/without STO buffer/capping layers. **b**, Temperature dependent resistivity for 6 uc LSMO films with different thickness of STO buffer layer. **c**, Kerr signal mapping of a patterned 6 uc LSMO film with size of 4.5×1 mm$^2$. The thicknesses of the STO buffer and capping layers are respectively 3 and 2 uc. **d**, Temperature dependent Kerr signal at region A ($S_2/L_6$), region B ($L_6$) and region C ($L_6/S_3$). The arrows indicate the Curie temperature ($T_C$).

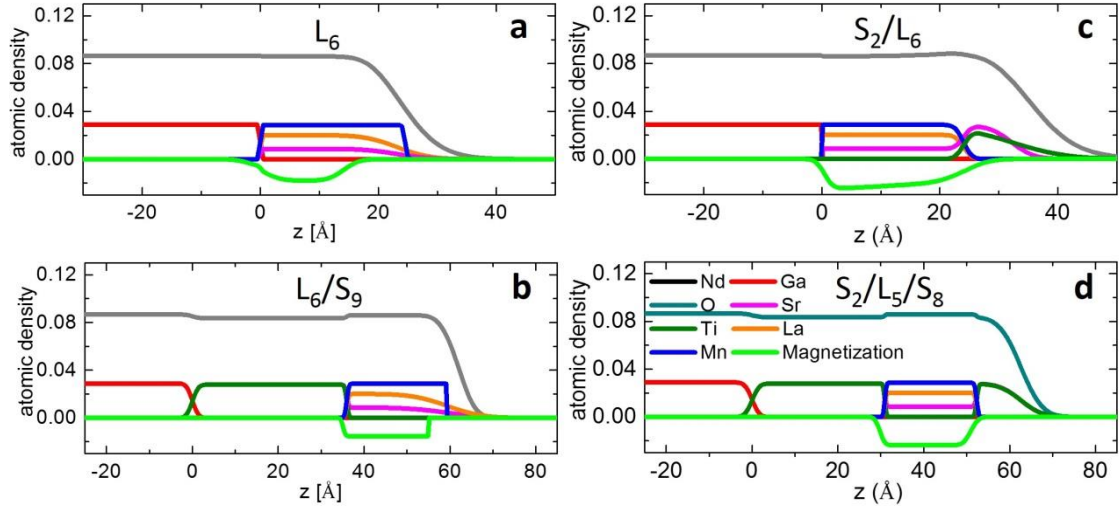

**Figure 3: Stoichiometry and magnetic profiles for different LSMO films with/without STO buffer/capping layers.** The depth profiles of the Ga, Ti, Mn, Nd, La, Sr, O atomic concentration and Mn magnetization were measured at 20 K. The detailed stacking of the samples: **a**, $L_6$; **b**, $L_6/S_9$; **c**, $S_2/L_6$; **d**, $S_2/L_5/S_8$.

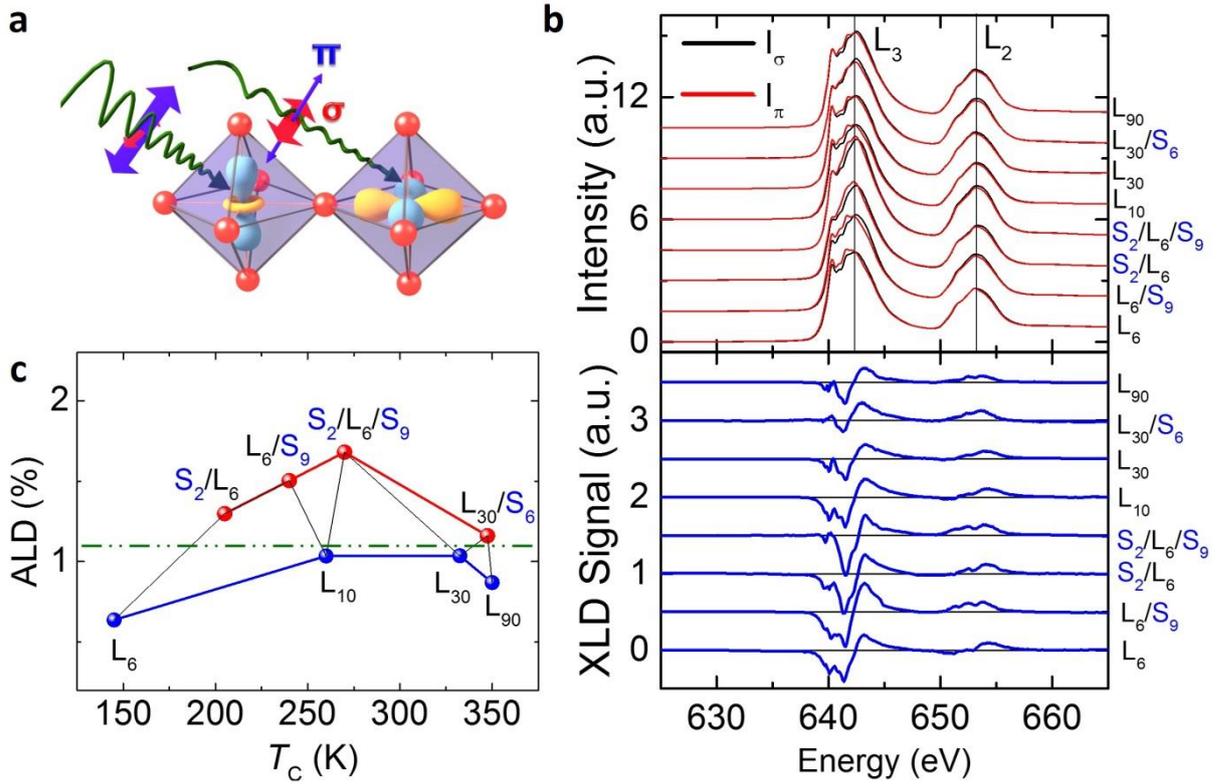

**Figure 4: Orbital reconstruction probed by X-ray absorption spectroscopy. a**, Sketch of the X-ray absorption measurement of $3z^2-r^2$ and $x^2-y^2$ orbital occupation by π- and σ-polarized lights, respectively. **b**, Polarized photons (in-plane σ, out of plane π) dependent XAS of Mn $L_{2,3}$ edge (top panel) and corresponding linear dichroism (bottom panel) of different LSMO films with/without STO buffer/capping layer. **c**, Integrated linear dichroism from 649-659 eV normalized by the $(I_\sigma+I_\pi)$ for LSMO films, ALD = $(I_\sigma-I_\pi)/(I_\sigma+I_\pi) \times 100\%$. All XLDs were measured at 300 K.

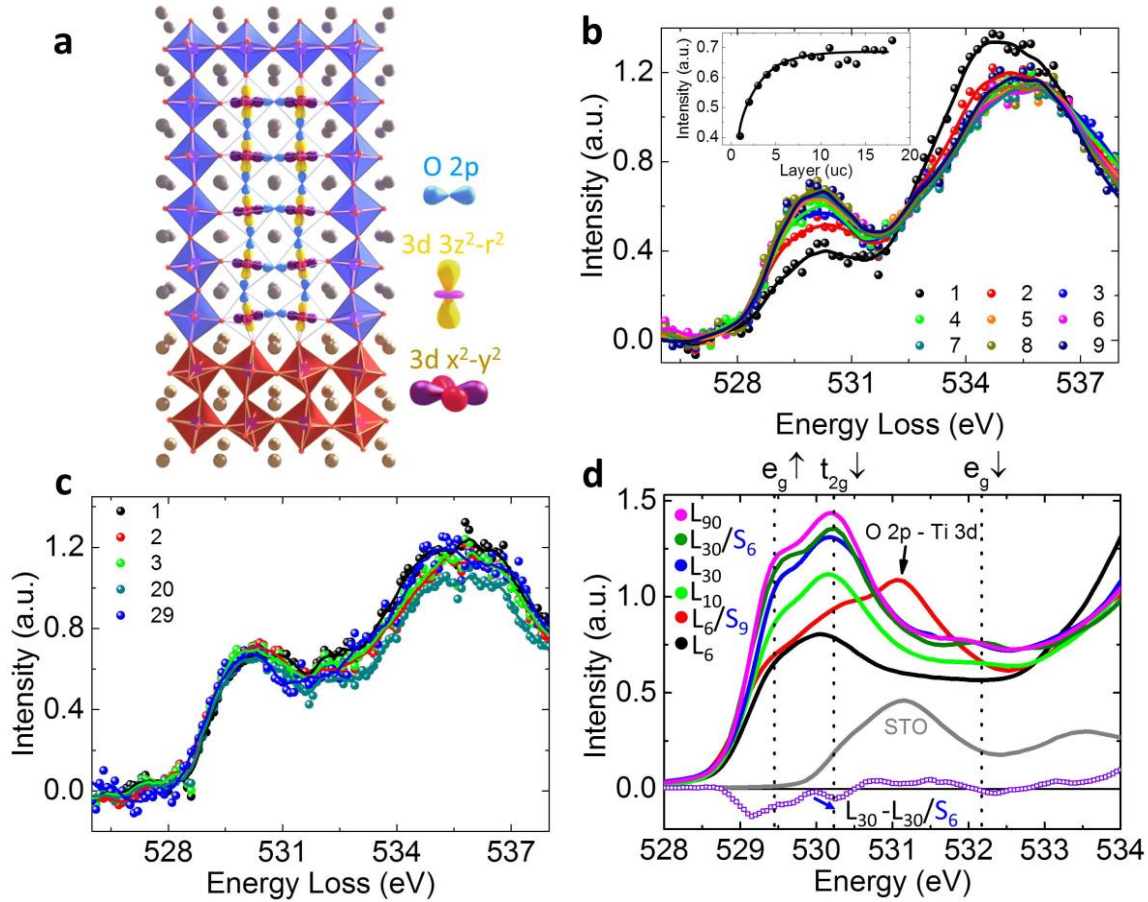

**Figure 5: Mn 3d and oxygen 2p orbital hybridization. a**, Schematic view of the evolution of the overlap of Mn-3d orbital with O-2p orbitals across the interface in the oxide heterostructure. Detailed O-K edge spectra of individual layers from EELS atomically-resolved mapping for a 30 uc LSMO ($L_{30}$) **b** and a buffered 30 uc LSMO ($L_{30}/S_6$) **c**. **d**, XAS O-K edge spectra of different LSMO films with/without STO buffer layer measured at 50 K. The grey curve is the O-K edge of a pure STO crystal. The open blue curve is intensity difference between $L_{30}$ and $L_{30}/S_6$.

# Supporting Information

**Thickness dependent properties in oxide heterostructures driven by structurally induced metal-oxygen hybridization variations**

Zhoaliang Liao, Nicolas Gauquelin, Robert J. Green, Sebastian Macke, Julie Gonnissen, Sean Thomas, Zhicheng Zhong, Lin Li, Liang Si, Sandra Van Aert, Philipp Hansmann, Karsten Held, Jing Xia, Johan Verbeeck, Gustaaf Van Tendeloo, George A. Sawatzky, Gertjan Koster, Mark Huijben*, Guus Rijnders

## 1. Measurement of B-O-B bond angle

Using statistical parameter estimation theory,[1-3] the 2D coordinates of each atomic column of the $La_{2/3}Sr_{1/3}MnO_3$ (LSMO) on $NdGaO_3$ (NGO) (110) heterostructure have been determined from the Annular Bright Field scanning transmission electron microscopy (ABF-STEM) image shown in Fig. S1a. Therefore, a parametric model in which projection images of the atomic columns are described using Gaussian peaks has been assumed. The parameters of this model, including the positions, height, and width of the intensity peaks, were determined using the least-squares estimator. The refined model evaluated at these estimated parameters is shown in Fig. S1b. From the estimated positions of the oxygen and B site (Ga or Mn) atomic columns, as indicated in the right top corner of Fig. S1b, the B-O-B bond angle has been calculated. The result for the averaged B-O-B angle in every atomic layer is shown in Fig. 1b in the main text.

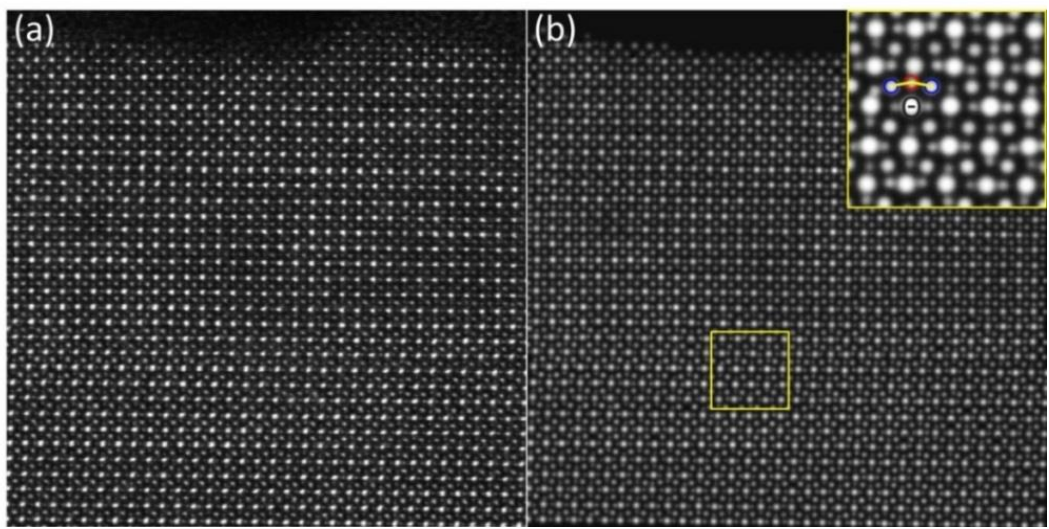

**Figure S1:** (a) Inversed ABF-STEM image of a 20 unit cells LSMO film on NGO (110) substrate. (b) Refined parametric model using statistical parameter estimation. Inset at right-top corner shows the zoom-in image of red square region.

## 2. Characterization of lattice parameters

The lattice constant of LSMO films was measured by PANalytical-X'Pert Materials Research Diffractometer (MRD) at high resolution mode. Figure S2a shows the θ-2θ scan of different thickness of LSMO with/without 6 unit cells (uc) SrTiO$_3$ (STO) buffer layers. The out-of-plane lattice constant $c$ is extracted from the position of LSMO (002)$_{pc}$ peak as indicated by the arrow. Here subscript pc represents pseudocubic indices. As shown in Fig. S2a, the 6 uc STO buffered 30 uc LSMO film ($L_{30}/S_6$) has lower peak position than non-buffered 30 uc LSMO ($L_{30}$), but is almost the same with 90 uc LSMO film ($L_{90}$). The unit cell lattice parameters ($a$, $b$, $c$) were further determined by reciprocal space mapping (RSM) of LSMO (240)$_{pc}$, (-240)$_{pc}$, (024)$_{pc}$, (0-24)$_{pc}$ peaks.[4] Here, we only shows (204)$_{pc}$ peaks as examples in Fig. S2b. Using the q$_z$ value of the (204)$_{pc}$ peak, the lattice constant $c$ can be determined from formula $c=4/q_z$. According to RSM, both the LSMO and STO layers are fully strained to NGO substrate, thus the LSMO and STO share the same in-plane lattice constant with NGO. RSM also illustrates that the $L_{30}/S_6$ film has lower q$_z$ and therefore bigger out-of-plane lattice constant ($c=4/q_z$) than non-buffered LSMO ($L_{30}$). In contrast to the STO buffer effect, the capping of 9 uc STO on top of LSMO ($S_9/L_{30}$) does not influence the LSMO lattice constant (see Fig. S2b). For thinner films, e.g., 15 uc LSMO film, it is hard to determine peak position in θ-2θ scan, but the $c$ can be measured from RSM. From the reciprocal space mapping, we can also determine the peak width and therefore estimate the uncertainty of the lattice parameter. According to the RSM and the θ-2θ scan, the lattice constant $c$ for different LSMO films and their uncertainties are obtained.

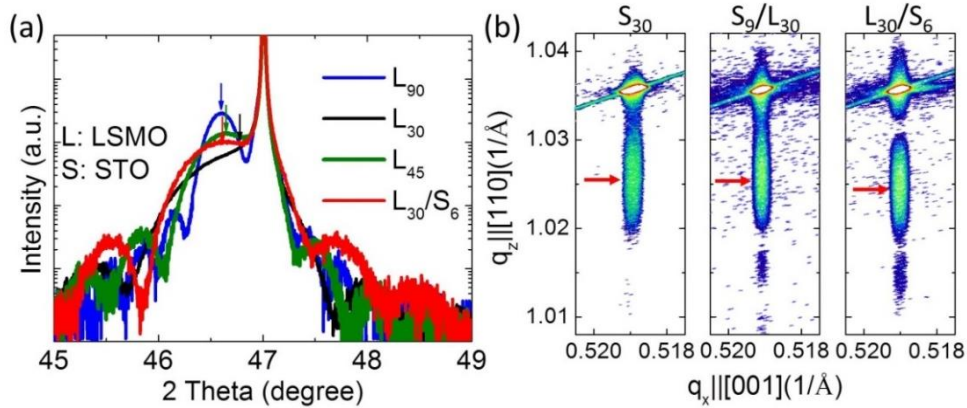

**Figure S2:** (a) θ-2θ scan of different thickness of LSMO with or without 6 uc STO buffer layers. The arrows indicate LSMO (002)$_{pc}$ peak positions. (b) RSM of (204)$_{pc}$ peaks of 30 uc LSMO films with/without STO buffer or capping layer. The red arrows indicate the LSMO peak positions. L and S represent LSMO and STO respectively, which are used for other figures as well.

## 3. The ABF-STEM image of STO buffered LSMO film

The ABF image of 6 uc STO buffered LSMO ($L_{30}/S_6$) film on NGO (110) substrate is shown in Fig. S3a-b. The tilt present in NGO quickly decays into the STO and is almost absent in the second STO unit cell already, similarly as previously observed for $L_6/S_9$.[5] The LSMO which is now connected to non-tilt STO layer presents no octahedral tilt. Figure S3b highlights the $BO_6$ octahedra across the LSMO/STO interface where the significantly reduced tilt angle is observed and compared with non-buffered LSMO film as showed in Fig. 1a in the main text.

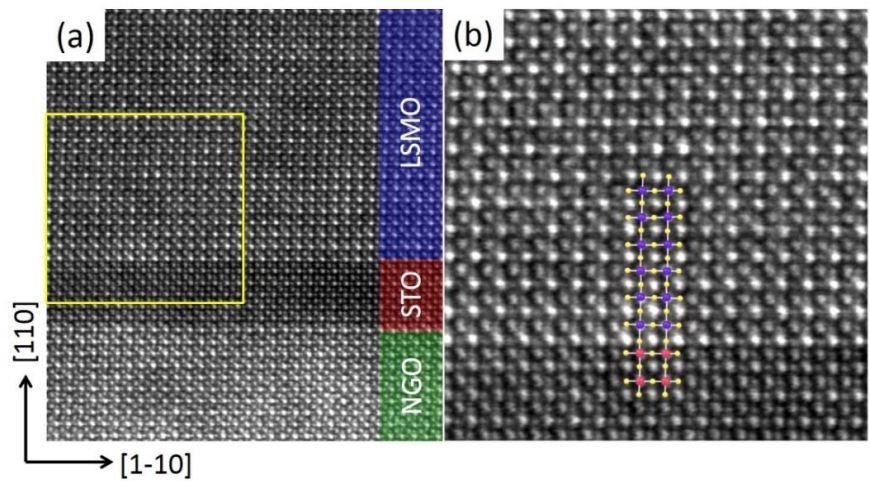

**Figure S3**: (a) ABF-STEM image of a 6 uc STO buffered 30 uc LSMO film. (b) Zoom-in of the yellow square region in (a). The colorized atom and bond highlight the octahedra across interface.

## 4. Patterned growth of STO and LSMO layers

By using shadow mask during growth, the film can be patterned as shown in Fig. S4. For the patterned LSMO sample, a shadow mask was first used to block one side of the NGO substrate during the growth of a STO buffer layer. Subsequently, a LSMO layer was deposited without shadow mask. The final step was to grow a STO capping layer with a shadow mask. This process produced three different regions—LSMO/NGO, LSMO/STO/NGO and STO/LSMO/NGO on the same substrate. The thicknesses of the STO and LSMO layers were monitored by reflection high energy electron diffraction (RHEED) intensity oscillations during growth.

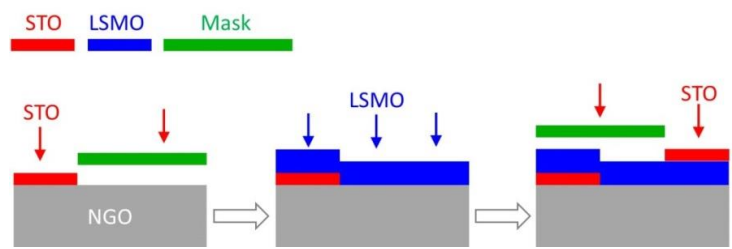

**Figure S4**: Process of patterned growth of STO and LSMO layers.

## 5. Determination of $T_C$ from Kerr measurement

To determine the Curie temperature ($T_C$), the patterned LSMO sample was cooled below Curie temperature using a liquid-helium-cooled cryostat in a magnetic field of 2000 gauss which will cant the magnetization along out-of-plane direction. The Kerr signal was measured by a Sagnac interferometer and the $T_C$ was determined from the second derivative of Kerr signal versus temperature as shown in Fig. S5. Since the magnetization easy axis is in-plane, applying a magnetic field to tilt the magnetization out of plane will have an effect of softening the paramagnetic to ferromagnetic phase transition, leading to less sharp transition in Kerr signal. However, the second derivative can largely amplify the transition and thus be used to determine the Curie temperature. As shown in Fig. S5, The $T_C$ is determined as the temperature point where the value of second derivate of Kerr signal reaches maximum negative value. There is a 1 K uncertainty in $T_C$ due to the signal noise associated with differentiating twice. Although this is an uncommon method of determining the $T_C$, but it still allows a quantitative comparison between the relative transition temperatures of different regions. Furthermore, the $T_C$ from Kerr measurement agrees well with the value measured by vibrating sample magnetometry (VSM), indicating a good estimation of $T_C$ using second derivate of Kerr signal.

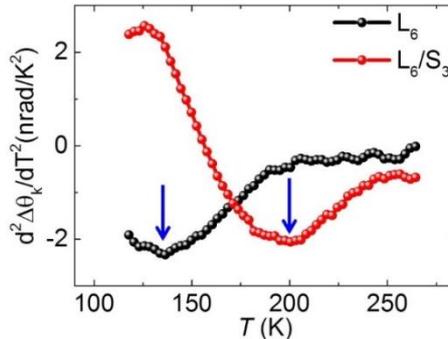

**Figure S5**: Second derivative of Kerr signal versus temperature for different regions of a patterned LSMO film mentioned in Fig. 2 in main text. Arrows indicate the Curie temperature.

## 6. Experimental and simulated RSXR Data

The chemical depth profiles of different LSMO samples have been determined from reflectivity curves measured at off-resonant energies, utilizing the optical contrast before and after each resonance. The film thickness, roughness, and a small contamination of light elements (such as oxygen and carbon) were taken as fit parameters, while the concentrations of the NGO, STO, and LSMO elements were fixed at stoichiometric values. Figure S6 shows the corresponding measurements and fits performed by the software ReMagX.[6,7] After determining the chemical profile, further measurements and modeling were used to determine the magneto-optical depth profile. For these measurements, a permanent magnet array producing a homogenous 0.6 Tesla field was inserted in the sample environment, aligning the magnetization in the film in-plane along the measurement scattering plane. Two different reflectivity curves at the Mn $L_3$ resonance were measured by using left $R_l$ and right circular $R_r$ polarized light. The bottom panel of each subfigure in Fig. S6 shows the asymmetry defined as $A = (R_l-R_r)/(R_l+R_r)$ and the corresponding fit. During fitting, the magnetic depth profile was assumed to be one homogeneous magnetic layer with in-plane magnetization and free thickness, position and magnetic roughness. As model inputs, the magneto optical constants were determined by the XMCD spectra taken from [8].

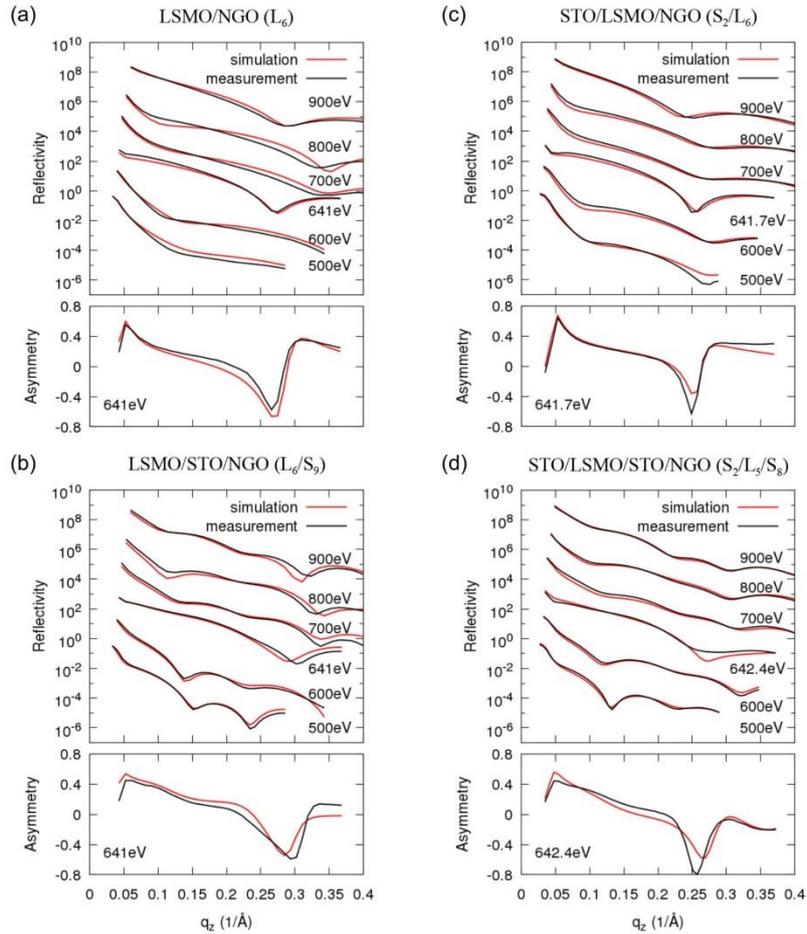

**Figure S6: Experimental and simulated RSXR data.** The data shown for the four samples correspond to the four extracted profiles in Fig. 3 of the main text. The upper panel for each sample contains $q_z$-scans at the specified energies using polarized light. The lower panel contains asymmetry spectra, obtained from the difference between left and right circular polarized light divided by the sum of both.

# 7. DFT calculation of structural effect on orbital hybridization

The impact of octahedral tilt on the electronic structure of LSMO is investigated using first principles density functional theory (DFT) calculations. The atomic relaxations are carried out with the VASP code[9,10] by using Projector augmented-wave (PAW) method.[11] Subsequently the electronic structure calculations were performed by using the WIEN2K code[12] with full-potential linearized augmented plane-wave method.[13] In both codes the generalized gradient approximation (GGA) of Perdew, Burke, and Ernzerhof (PBE)[14] functional was adopted to treat the exchange-correlation functions. The Sr doping in $LaMnO_3$ bulk ($La_{0.67}Sr_{0.33}MnO_3$) system is realized by using the virtual crystal approximation.

To study the effect of $MnO_6$ octahedral tilting, three different structures are considered in our study: cubic, $GdFeO_3$ (GFO)-type and NGO-type. For the cubic structure, there is no octahedral tilting, the Mn-O-Mn bond angle is 180°. The GFO- and NGO-type structures are those of LSMO but adopted to the GFO and NGO orthorhombic distortion, with NGO-type having larger octahedral tilting. In order to summarize the effects of the change of Mn-O-Mn bond angle and Mn-O bond length on the LSMO electronic structure, two group of calculations are carried out: In the first group (Group-I) we keep the Mn-O bond length constant. In this way any effect from Mn-O bond length is excluded, only the bond angle is changed (and subsequently the lattice parameters). In the second group (Group-II) we keep the Mn-Mn distance constant but otherwise relax the structure, rotating the $MnO_6$ octahedron in the xy plane and tilting them along the Z direction. This yields the GFO-type as an energy minimum; for the NGO-type we enhance the Mn-O-Mn angle according to the experimental NGO structure. Thus in the process of rotation and tilting the Mn-O bond length is also increased for Group-II calculations. By this confinement we can investigate the effect from both Mn-O-Mn bond angle and Mn-O bond length separately. All the structural parameters are summarized in Table S1. The half-metal ferromagnetic state of LSMO system is already obtained even without considering electron correlation, e.g. from DFT+U or DFT+DMFT methods. We also test the robustness of our results using a reasonably large interaction U in DFT+$U$ calculations: no remarkable change occurs. To include the effect of octahedral rotation and tilting, a √2×√2×2 supercell (20 atoms) was used in our electron structure calculations of the density of states (DOS) and bandwidth.

**Table S1: Summary of structural parameters (lattice constants, Mn-O-Mn angles and Mn-O bond length).** We consider three crystal structures (cubic, GFO- and NGO-type) and either keep the Mn-O bond length constant (Group-I) or the Mn-Mn distance constant of the crystal structure (Group-II).

| Structure | a(Å) | b(Å) | c(Å) | Mn-O-Mn angle θ (degree) | | | Mn-O bond (Å) | |
|---|---|---|---|---|---|---|---|---|
| | | | | $\theta_{xy}$ | $\theta_z$ | $<\theta>$ | xy | z |
| Group-I: Mn-O constant | | | | | | | | |
| Cubic | 3.848 | 3.848 | 3.848 | 180 | 180 | 180 | 1.924 | 1.924 |
| GFO-type | 3.834 | 3.793 | 3.810 | 162.8 | 162.8 | 162.8 | 1.924 | 1.924 |
| NGO-type | 3.829 | 3.633 | 3.771 | 151.6 | 151.2 | 151.5 | 1.924 | 1.924 |
| Group-II: Mn-Mn constant | | | | | | | | |
| Cubic | 3.848 | 3.848 | 3.848 | 180 | 180 | 180 | 1.924 | 1.924 |
| GFO-type | 3.848 | 3.848 | 3.848 | 159.4 | 160.5 | 159.9 | 1.952 | 1.953 |
| NGO-type | 3.848 | 3.848 | 3.848 | 151.6 | 151.2 | 151.5 | 1.984 | 1.987 |

Figure S7a shows an example of the orbital-resolved density of states (DOS) for cubic LSMO (θ = 180°) from which we can extract the bandwidth of the 3d majority $e_g$ and $t_{2g}$ orbitals, which is 4.83 eV and 3.336 eV, respectively. The DOS appearing at -7.5 eV to -4eV region arises from hybridization between oxygen 2p and Mn 3d orbitals and is excluded when calculation the bandwidth. By decreasing the bond-angle without changing the Mn-O bond length from 180° (cubic-type), 162.8° (GFO-type) to 151.5° (NGO-type) the effect of bond-angle on the electronic structure of LSMO (Group-I) is shown in Fig. S7b-c. There is a valley around -1 eV for the total DOS as shown in Fig. S7b, above and below which the DOS mainly comes from $e_g$ and $t_{2g}$, respectively. In cubic LSMO, the bandwidth of both $t_{2g}$ and $e_g$ are so big that the $t_{2g}$ and $e_g$ bands overlap, but their overlap is getting smaller with increasing the octahedral tilting (or said reducing bond angle) as shown in Fig. S7b. In NGO-type LSMO, the bandwidths of both $t_{2g}$ and $e_g$ become narrower. As a result a gap opens between $t_{2g}$ and $e_g$ orbital (zero DOS at the valley Fig. S7b). Figure S7c shows the partial DOS of majority $e_g$ orbital, clearly demonstrating the narrowing of the $e_g$ band with decreasing bond-angle.

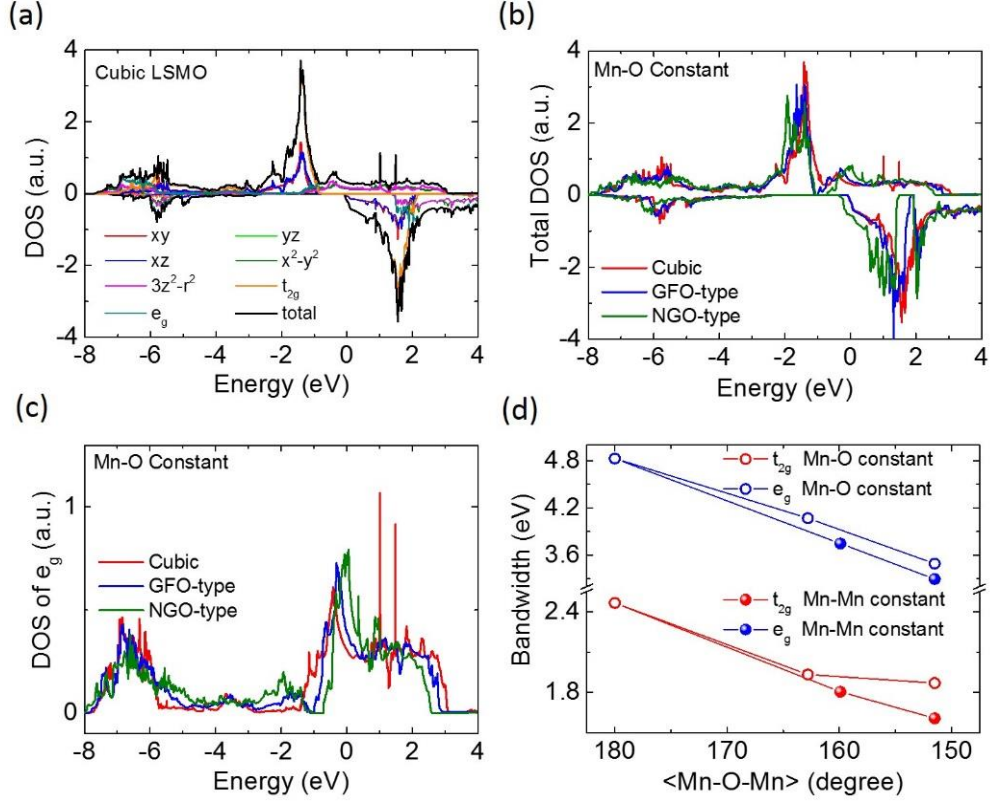

**Figure S7: Octahedral tilt effect on bandwidth.** (**a**) Calculated orbital- and spin-resolved DOS for both $e_g$ and $t_{2g}$ orbitals for cubic LSMO. (**b**) Spin-resolved total DOS and (**c**) majority $e_g$ DOS orbital of LSMO comparing cubic to GFO- and NGO-type of distortion with constant Mn-O bond length (Group-I). (**d**) Bandwidth as a function of tilting angle for both, constant Mn-O bond length (Group-I) and constant Mn-Mn distance (Group-II).

The bandwidth versus bond-angle is plotted in Fig. S7d. The bandwidths of both $t_{2g}$ and $e_g$ decrease with decreasing bond angle, e.g., in cubic LSMO the $e_g$ bandwidth is 4.83 eV, while it is 3.491 eV for NGO-type LSMO. Therefore, reducing the bond angle from 180° to 151.5° results in a 28% reduction of bandwidth. Since the oxygen 2p orbital points towards the $e_g$ orbital lobes not to those of the $t_{2g}$ orbitals, the $e_g$ orbitals have larger bandwidth as well as a larger the bandwidth reduction in Fig. S7d. Figure S7d also shows the calculated bandwidth of three different LSMO structures from Group-II where we keep the Mn-Mn distance constant but change the bond angle. The bandwidth is again reduced with decreasing bond angle, similar to our finding in Group-I LSMO. However, differently from Group-I, the bandwidth of Group-II LSMO gets narrower more quickly as the bond angle is reduced (see Fig. S7d). The bandwidth of $e_g$ is reduced by 32% when the bond angle changes from 180° to 151.5° for a constant Mn-Mn distance. That is, both the elongated Mn-O bond and the increased octahedral tilt reduce the bandwidth cooperatively. Therefore, the effect of octahedral tilt on bandwidth reduction will be enhanced somewhat if the Mn-Mn distance instead of the Mn-O bond length is locked. Taken together, both calculations show that the main driver behind the reduced bandwidth is the tilting of the Mn-O-Mn angle in the distorted structure.

The octahedral tilt will dominate the orbital hybridization in bulk LSMO, but the surface structure reconstruction will also play an important in determining the surface electronic structure of LSMO. To investigate the surficial structure relaxation, DFT calculation of the structure of 1 uc LSMO was performed. For simplicity, the $MnO_6$ octahedral tilt is eliminated, i.e., cubic structure is adopted in this calculation. By fixing the in-plane lattice constant while freeing the out-of-plane lattice and all atomic positions, the out-of-plane lattice constant $c_{film}^{relax}$ which is defined as Sr-Sr bond length is found to be 3.423 Å, only 88.9% of that the original non-relaxed unit cell (3.848 Å). This dramatic lattice compression is due to that the $(La_{2/3}Sr_{1/3}O)^{+0.67}/(MnO_2)^{-0.33}/(La_{2/3}Sr_{1/3}O)^{+0.67}$ system is polar and the intrinsic charge attractive force will push layers closer. Besides the Sr-Sr compression, the second feature of the film structure is that the distance between M and O at surface LaSrO layer is 2.07 Å, which is larger than the original 1.924 Å. The elongated M-O distance will dramatically reduce surfacial O-p to Mn-d hybridization.

The DFT calculation of 1 uc LSMO sandwiched between two STO layers which are both 2 uc thick was performed to investigate the impact of STO capping layer on the LSMO surface structure. It is found that the lattice constant $c$ of STO sandwiched LSMO film becomes 3.719 Å and the Mn-O distance is 1.934 Å, both of which are very close to the original bulk values of 3.848 Å and 1.904 Å. Our DFT calculation indicates that the STO capping layer can reduce the surface structure distortion, explaining the experimental observation of the recovery of surface magnetism due to the presence of STO capping layer as shown in Fig. 2 and Fig. 3 in main text.

## 8. Valence profile in LSMO films

The change of valence is another effect that can vary the Curie temperature. The X-ray absorption spectroscopy (XAS) as shown Fig. 4b of main text already indicates a nearly identical Mn valence in different LSMO films. The XAS also indicates a $Mn^{2+}$ peak located at ~ 640 eV. Regarding the fact that XAS detects near surface ~6 nm region, the presence of $Mn^{2+}$ in 90 uc ( = 35 nm) LSMO film then can only arise from surface region.[15,16] Such surficial $Mn^{2+}$ may also degrade the surface magnetism and contribute the surface dead layer.

The layer resolved valence profile is acquired from STEM electron energy loss spectroscopy (STEM-EELS). Figures S8a-b show the layer resolved EELS of Mn $L_{2,3}$ edge of $L_{30}$ and $L_{30}/S_6$ films, respectively. Very small shifts of the peaks are observed. For the LSMO/NGO sample, the interfacial Mn peak position is shifted a little to higher energy, suggesting a slightly higher Mn valence in relative to bulk region. For the STO buffered LSMO film ($L_{30}/S_6$), the peak position of interfacial layer moves a little bit to lower energy, suggesting slightly lower Mn valence compared to the bulk region.

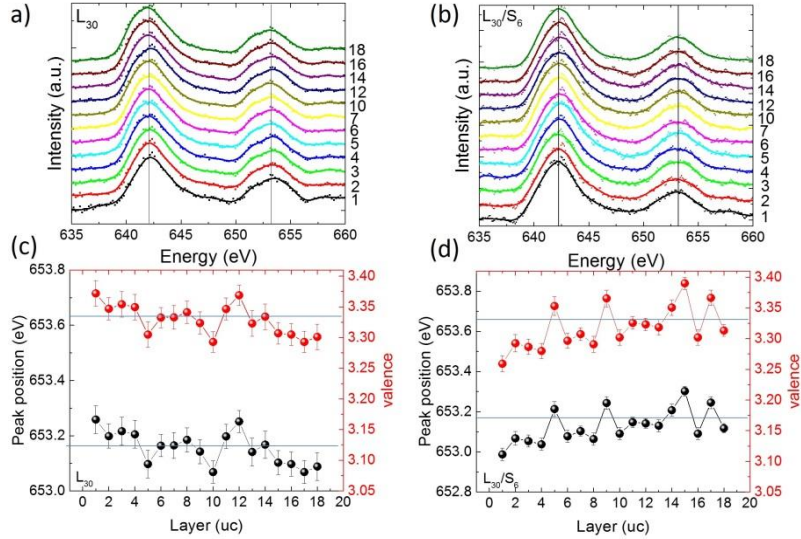

**Figure S8: Valence profile in LSMO films.** Layer resolved EELS of Mn $L_{2,3}$ peak across interface for (a) $L_{30}$ (a) and (b) $L_{30}/S_6$ and corresponding Mn $L_2$ peak position and estimated valence profile for (c) $L_{30}$ and (d) $L_{30}/S_6$.

According to previous reports by Mundy *et al.* [17] and Pellegrin *et al.* [18], the Mn $L_{2,3}$ peak position (*E*) is nearly linearly to Mn valence (*x*) and $t = \Delta x/\Delta E = \sim 0.42\ ev^{-1}$. With this slope *t*, the valence is estimated by the formula $x = 3.33+(E-E(3.33))\times t$ and resultant estimated valence profile is shown in Fig. S8c-d. The change of valence $\Delta x$ is smaller than 0.1 for both interfaces. The interfacial Mn layer has a little lower valence at LSMO/STO interface in comparison to a little higher valence at LSMO/NGO interface. Such trend agrees with the polar catastrophe model reported by Mundy et al. [17]. We would like to emphasize that the peak positions are more precisely determined from XAS which indicates nearly identical valence in different samples while the EELS has higher noise level which will lead to uncertainty of determining a precise peak position and thus may overestimate the change of valence. Although a more occupancy of Mn $e_g$ orbital will lead to a reduction of pre-edge intensity, but it is not significant that $I(Mn^{3+})/I(Mn^{4+}) = 0.8$.[19] Therefore, a much smaller change of pre-edge intensity will occur for a small change of valence and won't result to our observed pre-edge intensity profile. What is more, the Mn valence profile is expected to give rise to a little higher pre-edge intensity at the interface of LSMO/NGO, which is opposite to experimental observation that much lower intensity of pre-edge is observed for interfacial layer. While for LSMO/STO/NGO, a uniform pre-edge intensity is observed. Furthermore, local charge transfer won't play a role in thick film (e.g., $L_{30}$), hence a reassembled bulk like O-K edge and an increased $T_C$ due to STO buffer layer in 30 uc LSMO ($L_{30}/S_6$ vs. $L_{30}$) strongly suggest a central role of structure variation and resultant change of orbital hybridization in thickness driving phase transition.

## 9. Full spectra of O-K edge

XAS of oxygen K-edge was measured with the photon polarized along [001] axis of NGO. The full spectra of O-K edge for different LSMO films are shown in Fig. S9. The normalization of the spectra was done according to the total oxygen edge jump (i.e. scaled so that the pre- and post-edges were identical for all samples). This effectively normalizes to total oxygen content, allowing us to get a comparison between different samples. But for very thin films where the substrates start to make some contribution, it becomes very difficult to directly compare the intensity, but the line shapes still provide useful information for analyzing the characteristics of pre-edges, as discussed in main text.

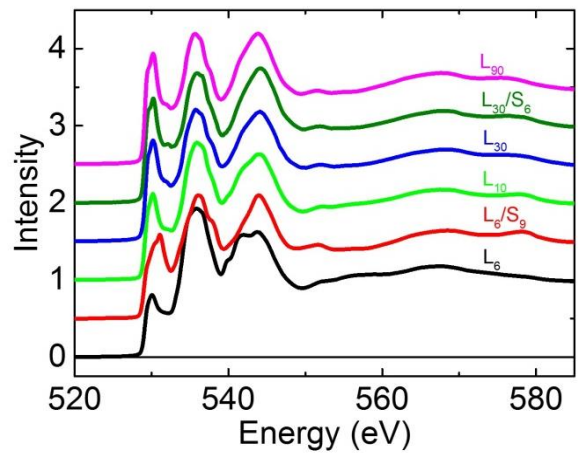

**Figure S9: Full spectra of O-k edge for different LSMO samples with/without STO buffer layers.**